\documentclass[Letter,11pt]{article}
\usepackage{graphicx,psfig} % para incluir eps
\begin{document}
\title{Quark-Antiquark Bound States in an Extended $QCD_2$
Model\footnote{\uppercase{T}alk given at 5th
Latin American Symposium on High Energy Physics \mbox{(V-SILAFAE)},
Lima Per{\'u}, \mbox{July 12-17, 2004.}}}

\date{}
\maketitle
\vspace{-2.1cm}
\begin{center}
{\large Pedro Labra{\~n}a, Jorge Alfaro}\\
\vspace{0.2cm}
Facultad de F\'{\i}sica,
Pontificia Universidad Cat\'olica de Chile. \\
%Vicu\~{n}a Mackena 4860, Macul, Santiago de Chile, Casilla 306, Chile.\\
plabrana@puc.cl, jalfaro@puc.cl\\
\medskip
\vspace{0.2cm}
{\large A.~A. Andrianov}\\
\vspace{0.2cm}
St.Petersburg State University and INFN, Sezione di Bologna.\\
andrianov@bo.infn.it
\end{center}

\begin{abstract}We study an extended QCD model in $2D$ obtained from QCD in $4D$
by compactifying two spatial dimensions and projecting onto the
zero-mode subspace. This system is found to induce a dynamical
mass for transverse gluons -- adjoint scalars in $QCD_2$, and to
undergo a chiral symmetry breaking with the full quark propagators
yielding non-tachyonic, dynamical quark masses, even in the chiral
limit. We construct the hadronic color singlet bound-state
scattering amplitudes and study quark-antiquark bound states which
can be classified in this model by their properties under Lorentz
transformations inherited from $4D$.
\end{abstract}

We study a QCD reduced model in $2D$ which can be
formally obtained from QCD in $4D$ by means of a classical
dimensional reduction from $4D$ to $2D$ and neglecting heavy K-K
(Kaluza-Klein) states. Thus only zero-modes in the harmonic
expansion in compactified dimensions are retained. As a
consequence, we obtain a two dimensional model with some
resemblances of the real theory in higher dimension, that is, in a
natural way adding boson matter in the adjoint representation to
$QCD_2$ \cite{light,Alfaro:2003yy}. The latter fields being scalars in $2D$ reproduce
transverse gluon effects \cite{adjoint}. Furthermore this model has a richer
spinor structure than just $QCD_2$ giving a better resolution of
scalar and vector states which can be classified by their
properties inherited from  $4D$ Lorentz transformations. The model
is analyzed in the light cone gauge and using large $N_c$ limit.
The contributions of the extra dimensions are controlled by the
radiatively induced masses of the scalar gluons as they carry a
piece of information of transverse degrees of freedom. We consider
their masses as large parameters in our approximations yet being
much less than the first massive K-K excitation. This
model might give more insights into the chiral symmetry breaking
regime of $QCD_4$. Namely, we are going to show that the inclusion
of  solely lightest K-K boson modes catalyze the generation of
quark dynamical mass and allows us  to overcome the problem of
tachyonic quarks present in $QCD_2$.

We start with the $QCD$ action in $(3+1)$ dimensions for one flavor
(extension to more flavors is straightforward):
%\vspace{-0.15cm}
%
\begin{equation}
S_{QCD}= \int d^4x \left[-\frac{1}{2{\tilde g}^2}tr(G_{\mu\nu}^2)
+ {\bar \Psi}\,(i\gamma^\mu \, D_\mu - m)\,\Psi \right].
\end{equation}

%\vspace{-0.2cm}

Follow the scheme of \cite{Alfaro:2003yy}
we proceed to make
a dimensional reduction of $QCD$, at the classical level, from $4D$
to $2D$. For this we consider the coordinates $x_{2,3}$ being
compactified in a 2-Torus, respectively the fields being periodic
on the intervals ($0\leq x_{2,3}\leq L=2\pi R$). Next we assume
$L$ to be small enough in order to get an effective model in $2D$
dimensions. Then by keeping only the zero
K-K modes, we get the following effective action in $2D$, after a
suitable rescaling of the fields:
%\vspace{-0.15cm}
%
\begin{eqnarray}
\label{Lmodelo2D}
S_2 &=& \int \!d^2x\,\,tr \!\left[-\frac{1}{2}F_{\mu\nu}^2+ (D_\mu
\phi_1)^2 + (D_\mu \phi_2)^2  \right] +
 {\bar \psi}_1\,(i\gamma^\mu \, D_\mu - m)\,\psi_1  \nonumber\\
&+& {\bar \psi}_2\,(i\gamma^\mu \, D_\mu - m)\,\psi_2 -
i\frac{g}{\sqrt{N_c}}\left({\bar \psi}_1\,\gamma^5\,\phi_1\,\psi_2 +
{\bar \psi}_2\,\gamma^5\,\phi_1\,\psi_1\right) \\
&-& i\frac{g}{\sqrt{N_c}}\left({\bar \psi}_1\,\gamma^5\,\phi_2\,\psi_1 -
{\bar \psi}_2\,\gamma^5\,\phi_2\,\psi_2\right) +
\frac{g^2}{N_c}\,tr[\phi_1,\phi_2]^2\,,
\nonumber
\end{eqnarray}
where we have defined the coupling constant of the model $g^2= N_c\,{\tilde g}^2/L^2$.
We expect \cite{Coleman1} the infrared mass generation for the two-dimensional
scalar gluons $\phi_i$. To estimate the masses
of scalar gluons $\phi_i$ we use the Schwinger-Dyson equations as
self-consistency conditions, we get:
%\vspace{-0.15cm}
%
\begin{equation}
 M^2=\frac{2 N_c {\tilde g}^2}{L^2} \int^\Lambda \!\!
\frac{d^2p}{(2\pi)^2} \,\frac{1}{p^2+M^2} \,
 = \frac{N_c {\tilde g}^2\,\Lambda^2}{8\pi^3}\,\,
 \log\frac{\Lambda^2+M^2}{M^2}\,, \label{glumass}
\end{equation}
%
%\vspace{-0.2cm}
%%%%%%%%%%%%%%%%%%%%%%%%%%%%%%%%%%%%%%%%%%%%%%%%%%%%%%%
\begin{figure}
\begin{center}
\includegraphics[width=0.9\textwidth]{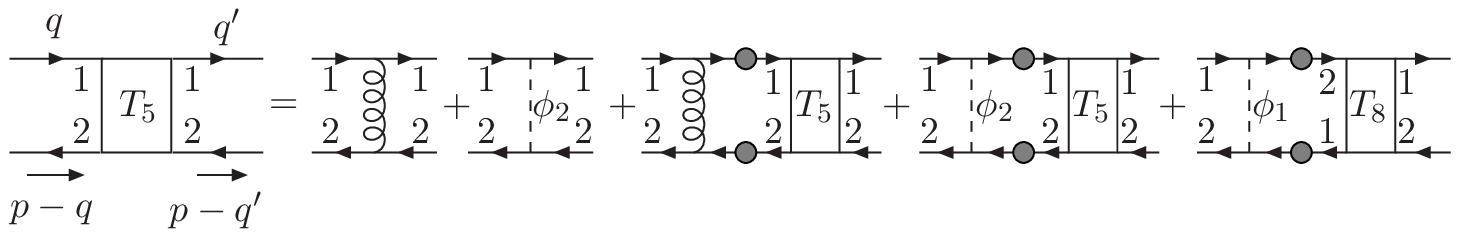}
\caption{Inhomogeneous Bethe-Salpeter equation for quark-antiquark
scattering amplitude $T_5$.} \label{F1}
\end{center}
\end{figure}
%%%%%%%%%%%%%%%%%%%%%%%%%%%%%%%%%%%%%%%%%%%%%%%%%%%%%%%
thus $M^2$ brings an infrared cutoff as expected. We notice that
the gluon mass remains finite in the large-$N_c$ limit if the QCD
coupling constant decreases as $1/N_c$ in line with the
perturbative law of $4D$ QCD. We adopt the
approximation $M \ll \Lambda\simeq 1/R$ to protect the low-energy
sector of the model and consider the momenta $|p_{0,1}| \sim M$.
Thereby we retain only leading terms in the expansion in
$p^2/\Lambda^2$ and $M^2/\Lambda^2$, and also
neglect the effects
of the heavy K-K modes in the low-energy
Wilson action. We observe that the limit $M \ll \Lambda$ supports
consistently both the fast decoupling of the heavy K-K modes and
moderate decoupling of scalar gluons \cite{Alfaro:2003yy}, the
latter giving an effective four-fermion interaction different from \cite{Burkardt}.
Also allow us to define the ``heavy-scalar''
expansion parameter $A=g^2/(2\pi\,M^2)=1/log \frac{\Lambda^2}{M^2} \ll 1$.
Now we proceed to the study of bound states of quark-antiquark. In
our reduction we have four possible combinations of quark
bilinears to describe these states with valence quarks:
$(\psi_1\,{\bar \psi}_1),(\psi_1\,{\bar \psi}_2),$ $
(\psi_2\,{\bar \psi}_1),(\psi_2\,{\bar \psi}_2 )$. We need
to compute the full quark-antiquark scattering amplitude
$T$ in the different channels. As an example we are going to show the
computing of $T_5$ which
correspond to the scattering ($q_1 + \bar{q}_2 \longrightarrow q_1 + \bar{q}_2$). It
satisfies, the equation given graphically in \mbox{Figure \ref{F1}}, in
the large $N_c$ limit and in ladder exchange approximation
(non-ladder contribution are estimated to be of higher order in
the $A$ expansion). Notice that in the equation for $T_5$ the amplitude
$T_8$ appears, which correspond to the process
$q_2 + \bar{q}_1 \longrightarrow q_1 + \bar{q}_2$. This means
that the equations for
$T_5$ and $T_8$ are coupled. In Figure \ref{F1} all internal
fermion lines correspond to dressed quark propagators which were determined
in \cite{Alfaro:2003yy} by using the large $N_c$ limit and
the one-boson exchange approximation. There are two kind of solutions for
the dressed quark propagators. In one solution, the perturbative one,
we have tachyonic quarks in the chiral limit as in $QCD_2$
and our model could be interpreted as a perturbation from the
result of chiral QCD  in $2D$. But we are not allowed to consider
that possibility because the spectrum for the lowest $q {\bar q}$
bound states becomes imaginary if one
takes into account the scalar field exchange. The second
solution, the non-perturbative one, supports non-tachyonic quarks with
masses going to zero, in the chiral limit and also yield real masses for
the $q {\bar q}$ bound states.
%%%%%%%%%%%%%%%%%%%%%% Fifura 2%%%%%%%%%%%%%%%%%%%%%%%%%%%%%%%%%
\begin{figure}
\begin{center}
\includegraphics[width=0.9\textwidth]{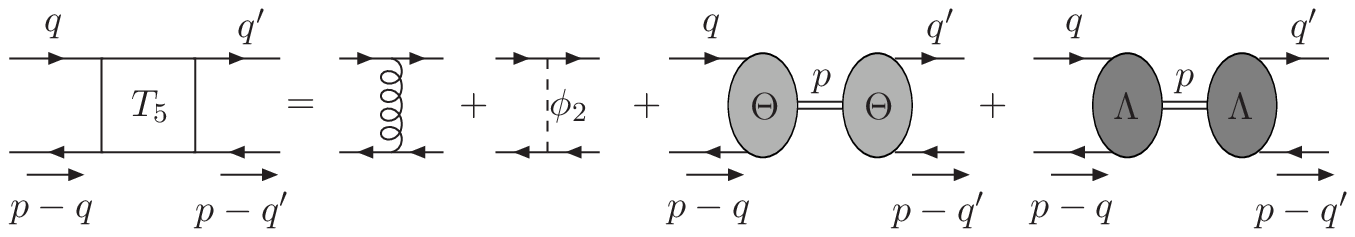}
\caption{$q\bar{q}$ scattering $T_5$ in the color-singlet
channel.} \label{F2}
\end{center}
\end{figure}
%%%%%%%%%%%%%%%%%%%%%%%%%%%%%%%%%%%%%%%%%%%%%%%%%%%%%%%
Having found the full quark propagator we proceed to
solve the inhomogeneous Bethe-Salpeter equation for $T_5$ and
$T_8$. We obtain for $T_5$ \mbox{(see Figure \ref{F2})}:
\begin{equation}
{\displaystyle T_5}\!
{\textstyle \stackrel{\alpha \beta, \gamma \delta ~~~~~~} {(q,q';p)}}
 =
-i\frac{g^2}{N}\,\frac{\gamma^{\alpha \delta}_-\,\gamma^{\beta
\gamma}_-}{(q-q')^2_-} - i\frac{g^2}{N}\,\frac{\sigma^{\alpha
\gamma}_3\,\sigma^{\beta \delta}_3}{[(q-q')^2 - M^2+i\epsilon]} \,+
\end{equation}
\[ +\, {\displaystyle \sum_j}\,\frac{i}{\big(p^2-\bar{m}^2_j + i\epsilon \big)}\,
\Theta^{\alpha\gamma}_j(q;p)\,\Theta^{\beta\delta}_j(q';p)
 +\, {\displaystyle \sum_k}\,\frac{i}{\big(p^2-\tilde{m}^2_k +i\epsilon \big)}\,
\Lambda^{\alpha\gamma}_k(q;p)\, \Lambda^{\beta\delta}_k(q';p)\, .
\]

There are no continuum states in the quark-antiquark amplitude--
only bound states at $p^2=\bar{m}^2$ and $p^2=\tilde{m}^2$, whose
residue yield the bound state wave functions $\Theta(p,r)$ and
$\Lambda(p,r)$. These bound states have a direct interpretation in
term of Dirac Bilinears of the theory in $4D$ \cite{Alfaro:2003yy}.
In particular, the pseudoscalar $4D$ states
are related with $\Lambda_k(r;p)$ which could be interpreted as
the vertex: (quark)-(antiquark)-($4D$ pseudoscalar meson). The
masses of the bound states and the vertex functions are fixed by
the solution of the homogeneous Bethe-Salpeter equation which, in
our model, yield to a eigenvalue problem, generalization of the
integral equation found by 't Hooft \cite{thooft1} in $QCD_2$. For
example the following eigenvalue integral equation determine the
mass spectrum of the pseudoscalar bound states (in the chiral
limit):
\begin{equation}
\label{n1} \tilde{m}^2 \, \phi(x) = \, -\frac{g^2}{\pi}\,
\int^{1}_{0} \! dy \, \frac{{\bf P}}{(x - y)^{2}}\,\, \phi(y) -\,
A\,\frac{\Sigma^{2}_{0}}{x^{1-\beta}} \int^{1}_{0} \! dy  \,
\frac{\phi(y)}{(1 - y)^{1-\beta}}
\end{equation}
\[
  - A\,\frac{\Sigma^{2}_{0}}{(1 - x)^{1-\beta}}
\int^{1}_{0} \! dy  \, \frac{\phi(y)}{y^{1-\beta}} +
A\,\frac{\Sigma^{2}_{0}}{[(1 - x)\,x]^{1-\beta}} \int^{1}_{0} \!
dy \, \phi(y) + A\,\Sigma^{2}_{0}\, \int^{1}_{0} \! dy  \,
\frac{\phi(y)}{[(1 - y)\,y]^{1-\beta}} ,
\]
where $\beta = A/2$.
To explore solutions to Eq.(\ref{n1}) we
examine small $A$. Evidently Eq.(\ref{n1}) does not mix even and
odd functions with respect to the symmetry $x \Longleftrightarrow
1 - x$. On the other hand the ground state should be an even
function. When inspecting the wave function end-point asymptotics
from the integral equation (\ref{n1}) one derives the following
even function as a  ground state solution for $A \rightarrow 0$
limit:
$\phi_0(x)=(4x[1-x])^{\frac{A}{2}}-\frac{1}{\pi}\,(4x[1-x])^{\frac{1}{2}}$.
This is basically a non-perturbative result that differs from 't
Hooft solution in the $A \rightarrow 0$ limit, giving $p^2=0$, as
we would expect from spontaneous chiral symmetry breaking in $4D$.
For the other massive states we are unable to find analytic
solutions, as happens with \mbox{'t Hooft} equation, but we could
estimate them working with the Hamiltonian matrix elements
$\tilde{m}^2(\phi,\phi)=(\phi,H\phi)$ and using the regular
perturbation theory, starting from \mbox{'t Hooft} solutions
($A=0$). In Table [1] we show lowest values of mass spectra for
scalar and pseudoscalar states.
\vspace{-0.5cm}

\section*{Discussions}
\vspace{-0.3cm}

Quantum Chromodynamics at low energies
has been decomposed by means of dimensional reduction from $4D$ to $2D$
and  a low energy effective model in $2D$ has
been derived.
In this model we did an
explicit analysis of meson bound states by solving the inhomogeneous and
the homogeneous Bethe-Salpeter equations, in the large $N_c$ limit and in the
ladder exchange approximation.
We found that the $2D$ model has four types of bound states which can be classified by their
properties under Lorentz transformations inherited from $4D$.
The $4D$ pseudoscalar and scalar sectors of the theory were
analyzed and in the quiral limit a massless solution for the pseudoscalar ground state
was found. We interpreted this solution as the ``pion'' of the
model.\\
Also, in solving the Bethe-Salpeter equations we found a solution to
the full quark propagators
yielding non-tachyonic dynamical quark masses, in contrast to what happen in
$QCD_2$.

\begin{table}[t]
\begin{center}
\begin{tabular}{ll} \hline \noalign{\smallskip}
Pseudoscalar & Scalar
\\\noalign{\smallskip} \hline \noalign{\smallskip}
%$p$ & ${\bar p}^2_1$ & ${\bar p}^2_1$ \\\hline
0& 974  \\
1286 & 1531\\
1741 & 1929 \\
2166 & 2356 \\\hline
\end{tabular}
\vspace{10pt}

Table [1]: Some values for the masses of bound states in $[MeV]$,\\
~~~~~~~~~~~where we have taken $\frac{g}{\sqrt{\pi}}=267 [MeV]$ and $A=0.22$.
\end{center}
%\vspace{-0.2cm}
\end{table}

\textbf{Acknowledgments:} P.L. thanks the organizers of the V-SILAFAE.
The work of A.A. is partially supported
by RFBR Grant and the Program ``Universities of
Russia: Basic Research''. The work of P.L. is
supported by a Conicyt Ph. D. fellowship (Beca Apoyo Tesis
Doctoral). The work of J.A. is partially supported by Fondecyt \#
1010967.

\end{document}